\documentstyle[epsbox,12pt]{article}

\begin{document}

\newcommand{\dfrac}[2]{\displaystyle{\frac{#1}{#2}}}

{\it University of Shizuoka}

\hspace*{9.5cm} {\bf US-98-04}\\[-.3in]

\hspace*{9.5cm} {\bf March 1998}\\[.3in]

\vspace*{.4in}

\begin{center}

{\large\bf New Aspects on a Seesaw Mass Matrix Model}\footnote{
Invited talk at the Workshop on {\it Fermion Mass and CP Violation},
Higashi-Hiroshima, Japan, March 5 - 6, 1998.} \\[.3in]

{\bf Yoshio Koide}\footnote{
E-mail: koide@u-shizuoka-ken.ac.jp} \\

Department of Physics, University of Shizuoka \\ 
52-1 Yada, Shizuoka 422-8526, Japan \\[.1in]

\vspace{.3in}

{\large\bf Abstract}\\[.1in]

\end{center}

\begin{quotation}
Recent development of the universal seesaw mass matrix model
is reviewed.
The model was proposed in order to explain why quark and 
lepton masses are so small compared with the electroweak scale 
$\Lambda_L =\langle\phi_L^0\rangle=174$ GeV. 
However, the recently observed top-quark mass $m_t\simeq 180$ GeV
seems to make an objection against the seesaw mass picture.
For this problem, it has recently pointed out that the seesaw 
mass matrix model is rather favorable to the fact 
$m_t\sim \Lambda_L$ if we consider the model with det$M_F=0$ 
for up-quark sector, where $M_F$ is a $3\times 3$ mass matrix 
of hypothetical heavy fermions $F$. 
The model can give a natural explanation why only top-quark 
acquire the mass of the order of $\Lambda_L$.
The model with det$M_U=0$ offers abundant new physics to us
(e.g., the fourth up-quark $t'$, FCNC, and so on).
\end{quotation}

\newpage

{\large\bf 1. Introduction}

\vglue.05in

One of  the most challenging problems in the particle 
physics is to give a unified understanding of quark and lepton
masses and mixings, of course, including the neutrino 
sector.
For this purpose, many models have been proposed [1].

In such a model-building, our interests are as follows:
Why is $m_t$ so extremely larger than $m_b$ 
in the third family in spite of $m_u \sim m_d$ 
in the first family ? 
Why is only $m_t$ of the order of $\Lambda_L$ (electroweak scale)?
Related to these topics, the recent development of 
the universal seesaw mass matrix model [2] and its 
special example are reviewed. 

As well-known, a would-be seesaw mass matrix for $(f, F)$ is
expressed as
$$
M = \left(\begin{array}{cc}
0 & m_L \\
m_R & M_F \\ 
\end{array} \right) = m_0 \left( 
\begin{array}{cc}
0 & Z_L \\
\kappa Z_R & \lambda Y_f \\
\end{array} \right) \ \ , 
\eqno(1.1)
$$
where $f=u, d, \nu, e$ are the conventional quarks and leptons,
$F=U, D, N, E$ are hypothetical heavy fermions, and 
they belong to $f_L = (2,1)$, $f_R = (1,2)$, 
$F_L = (1,1)$ and $F_R = (1,1)$ of 
SU(2)$_L \times $SU(2)$_R$.
The matrices $Z_L$, $Z_R$ and $Y_f$ are of the order one.
For the case $\lambda \gg \kappa \gg 1$, 
the mass matrix (1.1) leads to the well-known seesaw expression
$$
 M_f \simeq m_L M_F^{-1} m_R  \ .
\eqno(1.2)
$$

The mechanism  was first proposed [3] in order to answer the question 
of why neutrino masses are so invisibly small.
Then, in order to understand that the observed quark and lepton 
masses are considerably smaller than the electroweak scale,
the mechanism was applied to the quarks [2].

However, the observation of the top quark of 1994 [4] aroused
a question: Can the observed fact
$m_t \simeq 180$ GeV $\sim \Lambda_L = O(m_L)$ 
be accommodated to the universal seesaw mass matrix scenario?
Because $m_t\sim O(m_L)$ means $M_F^{-1}m_R \sim O(1)$.

For this question, a recent study gives the answer ``Yes":
Yes, we can do [5,6] by putting an additional constraint 
$$
{\rm det}M_F =0 \ .
\eqno(1.3)
$$
on the up-quark sector ($F=U$).
In the next section, we will review the mass generation scenario
on the basis of the universal seesaw mass matrix model with 
the constraint (1.3).

In Sec.~3, we discuss an abnormal structure of the quark 
mixing matrices and flavor changing neutral currents (FCNC) effects.
In Sec.~4, we review a model with  specific forms of the matrices
$Z_L$, $Z_R$ and $Y_f$, the so-called ``democratic seesaw mass 
matrix model" [5].
In Sec.~5, we give a short review of an application to the 
neutrino mass matrix.
Finally, Sec.~6 is devoted to the concluding remarks.

\vspace*{.2in}

{\large\bf 2. Energy scales and fermion masses }

\vglue.05in

For convenience, we take the diagonal basis of the matrix $M_F$.
Then, the condition (1.3) means that the heavy fermion mass matrix
$M_F$ in the up-quark sector is given by
$$ 
M_U=\lambda m_0 \left(
\begin{array}{ccc}
O(1) & 0 & 0 \\
0 & O(1) & 0 \\
0 & 0 & 0 
\end{array} \right) \ ,
\eqno(2.1)
$$
although the other heavy fermion mass matrices $M_F$ ($F\neq U$) 
are given by
$$
M_F=\lambda m_0 \left(
\begin{array}{ccc}
O(1) & 0 & 0 \\
0 & O(1) & 0 \\
0 & 0 & O(1) 
\end{array} \right) \ , \ \  (F\neq U) \ . 
\eqno(2.2)
$$ 
Note that for the third up-quark the seesaw mechanism does
not work (see Fig.~1).

\vspace*{.3in}
\begin{figure}[htb]

det$M_F\neq 0$ \ $\Longrightarrow$ \ Seesaw Mass
\hspace{17mm}
det$M_F = 0$ \ $\Longrightarrow$ \ Non-Seesaw Mass
\begin{picture}(650,160)(0,0)
\put(10,00){\thicklines\line(1,0){180}}
\put(100,00){\thicklines\line(0,1){120}}
\put(40,21){\thinlines\line(2,1){120}}
\put(40,19){\thinlines\line(2,1){120}}
\put(40,71){\thinlines\line(2,1){120}}
\put(40,69){\thinlines\line(2,1){120}}

\put(100,50){\circle*{5}}
\put(100,100){\circle*{5}}
\put(45,35){\circle{20}}
\put(40,32){$F_L$}
\put(40,82){{$F_R$}}
\put(45,85){\circle{20}}
\put(143,82){{$f_R$}}
\put(148,86){\circle{20}}
\put(143,132){{$f_L$}}
\put(148,136){\circle{20}}
\put(25,35){\thicklines\line(0,1){50}}
\put(25,35){\thicklines\line(1,0){8}}
\put(25,85){\thicklines\line(1,0){8}}
\put(5,53){{$M_F$}}
\put(105,35){{$\Lambda_R$}}
\put(105,85){{$\Lambda_L$}}
\put(105,5){{$\Lambda_S$}}
\put(145,10){{$m(F_L,F_R)\sim\Lambda_S$}}
\put(141,110){{$m(f_L,f_R)\sim\frac{\Lambda_L\Lambda_R}{\Lambda_S}$}}

\put(210,00){\thicklines\line(1,0){180}}
\put(300,00){\thicklines\line(0,1){120}}
\put(235,51){\thinlines\line(1,0){130}}
\put(235,49){\thinlines\line(1,0){130}}
\put(235,101){\thinlines\line(1,0){130}}
\put(235,99){\thinlines\line(1,0){130}}
\put(300,50){\circle*{5}}
\put(300,100){\circle*{5}}
\put(237,55){$F_{L3}$}
\put(245,59){\circle{21}}
\put(237,105){{$F_{R3}$}}
\put(245,109){\circle{21}}
\put(355,55){{$f_{R3}$}}
\put(362,59){\circle{21}}
\put(355,105){{$f_{L3}$}}
\put(362,109){\circle{21}}
\put(305,35){{$\Lambda_R$}}
\put(305,85){{$\Lambda_L$}}
\put(305,5){{$\Lambda_S$}}
\put(345,30){{$m(F_{L3},f_{R3})\sim\Lambda_R$}}
\put(345,80){{$m(f_{L3},R_{R3})\sim\Lambda_L$}}
\end{picture}

\vglue.1in

\centerline{Fig.~1. Seesaw and non-seesaw masses}
\end{figure}

\vspace*{.2in}

Therefore, the mass generation at each energy scale is as follows.
First, at the energy scale $\mu=\Lambda_S$, the heavy fermions $F$,
except for $U_3$, acquire the masses of the order of $\Lambda_S$.
Second, at the energy scale $\mu=\Lambda_R$, the SU(2)$_R$ 
symmetry is broken, and the fermion $u_{R3}$ generates a mass
term of the order of $\Lambda_R$ by pairing with $U_{L3}$.
Finally, at $\mu=\Lambda_L$, the SU(2)$_L$ symmetry is broken,
and the fermion $u_{L3}$ generates a mass term of the order 
$\Lambda_L$ by pairing with $U_{R3}$. 
The other fermions $f$ acquire the well-known seesaw masses
(1.2).
The scenario is summarized in Table 1.

\vspace{.2in}
\centerline{Table 1. Fermion mass generation scenario}
\vglue.05in
\begin{tabular}{|c|c|c|}\hline
Energy scale &  $d$- \& $e$-sectors &
$u$-sector ($i\neq 3$) \\ \hline
{\large At} $\mu=\Lambda_S\sim \lambda m_0$ & $m(F_L, F_R) \sim \Lambda_S$  
& $m(U_{Li}, U_{Ri}) \sim \Lambda_S$ \\ \hline
{\large At} $\mu=\Lambda_R\sim \kappa m_0$ &  & $m(u_{R3}, U_{L3}) 
\sim \Lambda_R$  \\ \hline 
{\large At} $\mu=\Lambda_L\sim m_0$ & 
& $m(u_{L3}, U_{R3})\sim \Lambda_L$  \\[.2in] 
 & $m(f_L, f_R)\sim \displaystyle\frac{\Lambda_L \Lambda_R}{\Lambda_S}$  & 
$m(u_{Li}, u_{Ri}) \sim \displaystyle\frac{\Lambda_L \Lambda_R}{\Lambda_S}$ 
 \\ \hline
\end{tabular}

\vglue.1in

Thus, we can understand why only top quark $t$ acquires  
the mass $m_t \sim O(m_L)$.

\vspace{.2in}

\centerline{Table 2. Neutrino mass generation scenarios}
\vglue.05in

\begin{tabular}{|c|c|c|}\hline
Energy scale &  Scenario A & Scenario B  \\ \hline
{\large At} $\mu=\Lambda_{SS}$ & $m(N_L, N_R) \sim \Lambda_{SS}$  & \\ \hline
{\large At} $\mu=\Lambda_S$ &  & $m(N_+, N_+^c) \sim \Lambda_S$ \\ \hline
{\large At} $\mu=\Lambda_R$ &  & $m(\nu_R, N_-)\sim \Lambda_R$  \\ \hline 
{\large At} $\mu=\Lambda_L$ & $m(\nu_L, \nu_R)\sim 
\displaystyle\frac{\Lambda_L \Lambda_R}{\Lambda_{SS}}$ &
$m(\nu_L, \nu_L^c) \sim \displaystyle\frac{\Lambda_L^2}{\Lambda_S}$  \\ \hline
\end{tabular}

\vglue.05in
\hfill $N_\pm =(N_L \pm N_R^c)/\sqrt{2}$

\vspace{.2in}
Next, we discuss the neutrino mass generation.
At present, we have two scenarios.

One (Scenario A) is a trivial extension of 
the present model:
we introduce a further large energy scale 
$\Lambda_{SS}$ in addition to $\Lambda_S$, and
we assume that $M_F \sim \Lambda_S$ ($F=U,D,E$), while 
$M_N \sim \Lambda_{SS}$ ($\Lambda_{SS}\gg \Lambda_S$)

Another scenario (Scenario B) is more attractive 
because we does not introduce an additional energy scale.
The neutral heavy leptons are singlets of 
SU(2)$_L\times$SU(2)$_R$ and they do not have
U(1)-charge. 
Therefore, it is likely that they acquire 
Majorana masses $M_M$ together with the Dirac 
masses $M_D\equiv M_N$ at $\mu=\Lambda_S$.
Then, the conventional light neutrino masses
$m_\nu$ are given with the order of 
$$
m_\nu \sim \frac{\Lambda_L^2}{\Lambda_S}=
\frac{1}{\kappa}\frac{\Lambda_L\Lambda_R}{\Lambda_S} \ .
\eqno(2.3)
$$
In order to explain the smallness of $m_\nu$, 
the model [7,8]
requires that the scale $\Lambda_R$ must be extremely 
larger than $\Lambda_L$.

On the other hand, the scenario A allows a case with
a lower value of $\Lambda_R$.
Then, we may expect abundant new physics effects as 
we discuss later.

The neutrino mass generation scenarios are summarized
in Table 2.

\vspace{.2in}

{\large\bf 3. Abnormal structure of $U_R^u$ and FCNC}

\vglue.05in

The most excited features of the present model is an 
abnormal structure of the right-handed fermion mixing
matrix $U_R$ [9].

For the down-quark sector, where the seesaw expression
(1.2) is valid, the mixing matrices $U_L^d$ and $U_R^d$ 
are given by
$$ 
U_L^d = \left(
\begin{array}{cc}
A_d & \frac{1}{\lambda} C_d \\
\frac{1}{\lambda} C'_d & B_d 
\end{array} \right) \ , \ \ 
U_R^d \simeq \left(
\begin{array}{cc}
A_d^* & \frac{\kappa}{\lambda} C_d^* \\
\frac{\kappa}{\lambda} C^{\prime \ast}_d & B_d 
\end{array} \right) \ , 
\eqno(3.1)
$$
where $A, B, C \sim O(1)$.
However, in contrast with the down-quark sector, 
for the up-quark sector, 
where the seesaw expression is not valid any more,
the mixing matrices $U_L^u$ and $U_R^u$ 
are given by
$$
U_L^u = \left(
\begin{array}{lll|lll} 
\ast & \ast & \ast & \sim\frac{1}{\lambda} & 
\sim\frac{1}{\lambda} & \sim\frac{1}{\lambda} \\
\ast & \ast & \ast & \sim\frac{1}{\lambda} 
& \sim\frac{1}{\lambda} & \sim\frac{1}{\lambda} \\
\ast & \ast & \ast & \sim\frac{1}{\lambda} & 
\sim\frac{1}{\lambda} & \sim\frac{1}{\lambda} \\ 
\hline
\sim\frac{1}{\lambda} & \sim\frac{1}{\lambda} 
& \sim\frac{1}{\lambda} & \ast & \ast & \ast \\
\sim\frac{1}{\lambda} & \sim\frac{1}{\lambda} 
& \sim\frac{1}{\lambda} & \ast & \ast & \ast \\
\sim\frac{1}{\lambda} & \sim\frac{1}{\lambda} 
& \sim\frac{1}{\lambda} & \ast & \ast & \ast 
\end{array} \right) \ ,
\eqno(3.2)
$$
$$
U_R^u = \left(
\begin{array}{lll|lll} 
\ast & \ast & \ast & \sim\frac{\kappa}{\lambda} & 
\sim\frac{\kappa}{\lambda} 
& \sim\frac{\kappa}{\lambda} \\
\ast & \ast & \ast & \sim\frac{\kappa}{\lambda} 
& \sim\frac{\kappa}{\lambda} & \sim \frac{\kappa}{\lambda} \\
\sim\frac{\kappa}{\lambda} & \sim\frac{\kappa}{\lambda} 
& \sim\frac{\kappa}{\lambda} & \ast & \ast & \ast \\
\hline
\ast & \ast & \ast & \sim\frac{\kappa}{\lambda} 
& \sim\frac{\kappa}{\lambda} & \sim\frac{\kappa}{\lambda} \\ 
\sim\frac{\kappa}{\lambda} & \sim\frac{\kappa}{\lambda} 
& \sim\frac{\kappa}{\lambda} & \ast & \ast & \ast \\
\sim\frac{\kappa}{\lambda} & \sim\frac{\kappa}{\lambda} 
& \sim\frac{\kappa}{\lambda} & \ast & \ast & \ast 
\end{array} \right) \ ,
\eqno(3.3)
$$
where the symbol $\ast$ denotes numerical factors of $O(1)$.
Note that the right-handed up-quark mixing matrix $U_R^u$
has a peculiar structure as if the third and fourth rows
of $U_R^u$ are exchanged each other in contrast to 
$U_L^u$.

This is understood from the following expression of 
the Hermitian matrixes $H_L$ and $H_R$:
On  the diagonal basis of $M_F$, the Hermitian matrices 
for the up-quark sector are given by
$$
\begin{array}{ll}
H_L^u = M M^\dagger & H_R^u = M^\dagger M \\
= 
m_0^2 \left(\begin{array}{cc}
Z_L Z_L^\dagger & \lambda Z_L Y_u^\dagger \\ 
\lambda Y_u Z_L^\dagger & 
\lambda^2 Y_u Y_u^\dagger + \kappa^2 Z_R Z_R^\dagger \\ 
\end{array} \right)  \ & 
= m_0^2 
\left(\begin{array}{cc}
\kappa^2 Z_R^\dagger Z_R & \kappa \lambda Z_R^\dagger Y_u \\ 
\kappa\lambda Y_u^\dagger Z_R & 
\lambda^2 Y_u^\dagger Y_u + Z_L^\dagger Z_L \\ 
\end{array} \right)   \ \\
=\left(
\begin{array}{ccc|ccc}
\ast & * & * & 0 & * \lambda & * \lambda \\
\ast & * & * & 0 & * \lambda & * \lambda \\
\ast & * & * & 0 & * \lambda & * \lambda  \\
\hline
0 & 0 & 0 & * \kappa^2 & * \kappa^2 & * \kappa^2 \\
\ast  \lambda & * \lambda & * \lambda & 
* \kappa^2 & * \lambda^2 & * \kappa^2 \\
\ast  \lambda & * \lambda & * \lambda & 
* \kappa^2 & * \kappa & * \lambda^2  \\
\end{array} \right) 
& 
=\left(
\begin{array}{ccc|ccc}
\ast \kappa^2 & * \kappa^2 & * \kappa^2 & 
0 & * \kappa\lambda & * \kappa\lambda  \\
\ast \kappa^2 & * \kappa^2 & * \kappa^2 & 
0 & * \kappa\lambda & * \kappa\lambda  \\
\ast \kappa^2 & * \kappa^2 & * \kappa^2 & 
0 & * \kappa\lambda & * \kappa\lambda  \\
\hline
0 & 0 & 0 & *  & *  & *  \\
\ast \kappa \lambda & * \kappa\lambda & * \kappa\lambda & 
*  & * \lambda^2 & *  \\
\ast \kappa \lambda & * \kappa\lambda & * \kappa\lambda & 
*  & * & * \lambda^2 \\
\end{array} \right) \ .
\end{array}
\eqno(3.4)
$$
That is, in the present model, the roles of $u_{3R}$ and $U_{1R}$
are exchanged each other in $H_R^u$. 
This means that the mass-partners  are given by
$$
\begin{array}{ll}
u \simeq u_1 = (u_{1L}, u_{1R}) \ , & d \simeq d_1 = (d_{1L}, d_{1R}) \ ,  \\
c \simeq u_2 = (u_{2L}, u_{2R}) \ , & s \simeq d_2 = (d_{2L}, d_{2R}) \ ,  \\
t \simeq u_3 = (u_{3L}, U_{1R}) \ , & b \simeq d_3 = (d_{3L}, d_{3R}) \ ,  \\
t' \simeq u_4 = (U_{1L}, u_{3R}) \ , & D \simeq d_4 = (D_{1L}, D_{1R}) \ ,  \\
C \simeq u_5 = (U_{2L}, U_{2R}) \ , & S \simeq d_5 = (D_{2L}, D_{2R}) \ ,  \\
T \simeq u_6 = (U_{3L}, U_{3R}) \ , & B \simeq d_6 = (D_{3L}, D_{3R}) \ ,  \\
\end{array} \eqno(3.5)
$$
where, for convenience, the numbering of the heavy up-quarks $U$ has been
changed from the definition based on (2.1).

As seen in (3.5), for a model with a low value of 
$\Lambda_R$ (for example, $\kappa \sim 10$), 
we may expect [9]
a single production of $t'$ with 
$m_{t'}\simeq \kappa m_t \sim$ a few TeV, 
through the exchange of 
$W_R$ : $ d+u \rightarrow t' +d$, i.e.,
$p +p \rightarrow t' +X$ at LHC, 

In the present model, the FCNC effects induced by 
the abnormal structure of the mixing matrix appear.
The magnitudes are  proportional to the factor
$$
\xi^f = U_{fF} U_{fF}^\dagger \ , 
\eqno(3.6)
$$
where
$$ 
U=\left( 
\begin{array}{cc}
U_{ff} & U_{fF} \\
U_{Ff} & U_{FF} 
\end{array} \right) \ . 
\eqno(3.7)
$$
We can obtain sizable values of 
$|(\xi_R^u)_{tc}|$ and $|(\xi_R^u)_{tu}|$,  
although the other factors are invisibly small, e.g., 
$|(\xi_R^u)_{ij}|\sim (\kappa/\lambda)^2$, \  
$|(\xi_L^u)_{ij}|\sim (1/\lambda)^2$.  
Therefore, if $\kappa\sim 10$, the FCNC effects appear visibly 
in the modes related to top-quark.  
Then, for example, we may expect the following
single-top-production: 
$ e^- + p\rightarrow e^- +t +X $ at HERA, 
$e^- +e^+ \rightarrow t +\overline{c}$ at JLC,
and so on.

The numerical results for a model with a specific matrix 
form can be found in Ref.[9].

\vspace{.2in}

{\large\bf 4. Democratic seesaw mass matrix model}

\vglue.05in

So far, we have not assumed explicit structures of the 
matrices $Z_L$, $Z_R$ and $Y_f$.
Here, in order to give a realistic numerical example,
we put the following working hypotheses [5]: 

\noindent
(i) The matrices $Z_L$ and $Z_R$, which are universal
for quarks and leptons, have the same 
structure:
$$
Z_L = Z_R \equiv Z = {\rm diag} (z_1, z_2, z_3) \ \ , 
\eqno(4.1)
$$
with $z_1^2 + z_2^2 + z_3^2 = 1$, 
where, for convenience, we have taken a basis on which 
the matrix $Z$ is diagonal. 

\noindent
(ii) The matrices $Y_f$, which have structures 
dependent on the fermion sector $f=u,d,\nu,e$, take
a simple form [(unit matrix)+(a rank one matrix)]:
$$
Y_f = {\bf 1} + 3 b_f X \ \ . 
\eqno(4.2)
$$
(iii) The rank one matrix is  given by
a democratic form
$$
X = \frac{1}{3}\left(\begin{array}{ccc}
1 & 1 & 1 \\
1 & 1 & 1 \\
1 & 1 & 1 \\
\end{array} \right) \  , 
\eqno(4.3)
$$
on the family-basis where the matrix $Z$ is diagonal.

\noindent
(iv) In order to fix the parameters $z_i$, we 
tentatively take $b_e = 0$ for the charged lepton sector,
so that the parameters $z_i$ are given by
$$
\frac{z_1}{\sqrt{m_e}} = \frac{z_2}{\sqrt{m_\mu}} = 
\frac{z_3}{\sqrt{m_\tau}} = \frac{1}{\sqrt{m_e + m_\mu + m_\tau}} \ \ . 
\eqno(4.4)
$$

The mass spectra are essentially characterized by the parameter $b_f$.
The fermion masses $m_i^f$ versus $b_f$ are illustrated in Fig.~2.
At $b_f=0$, the charged lepton masses have been used as input values
for the parameters $z_i$.
Note that at $b_f=-1/3$, the third fermion mass takes a maximal 
value, which is independent of $\kappa/\lambda$.
Also note that at $b_f=-1/2$ and $b_f=-1$, two fermion masses 
degenerate.

\begin{minipage}[tl]{10cm}
\epsfile{file=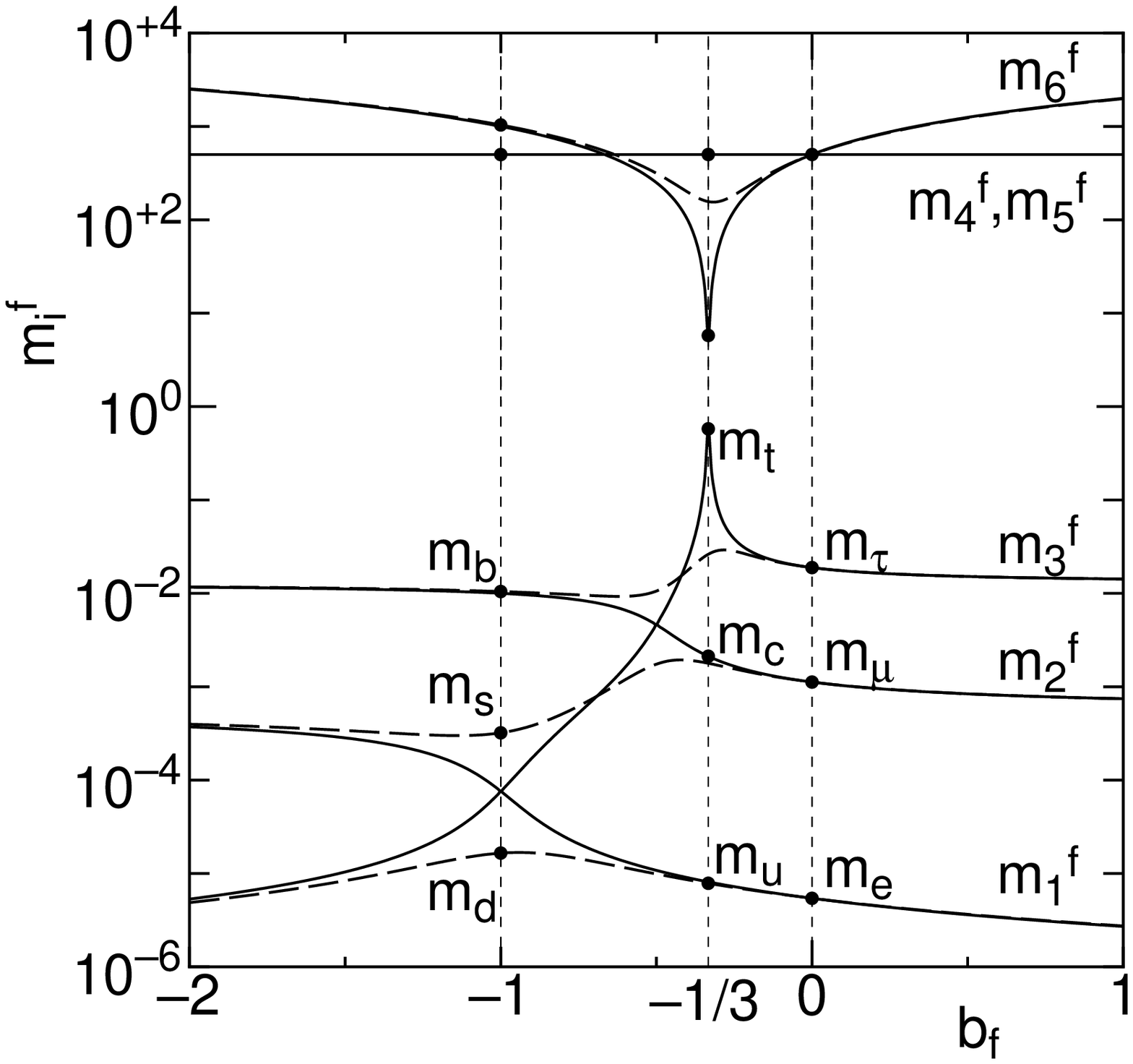,scale=0.45}
\end{minipage}
\begin{minipage}[tr]{5.5cm}
{\small Fig.~2. Masses $m_i$ ($i=1,2,\cdots,6$) versus $b_f$ for the 
case $\kappa=10$ and $\kappa/\lambda=0.02$.
The solid and broken lines represent the cases arg$b_f=0$ and 
arg$b_f=18^\circ$, respectively. The figure was quoted from Ref.~[12].
}
\end{minipage}

We take $b_u = -1/3$  for up-quark sector, 
because, at $b_u=-1/3$, we can obtain the maximal top-quark mass 
enhancement (see Fig.~2)
$$
m_t \simeq \frac{1}{\sqrt{3}} m_0 \ ,
\eqno(4.5)
$$
and  a successful relation
$$
 \frac{m_u}{m_c} \simeq \frac{3}{4}\frac{m_e}{m_\mu}  \ , 
\eqno(4.6)
$$
independently of the value of $\kappa/\lambda$.

The value of $\kappa/\lambda$ is determine from the observed ratio
$m_c/m_t$ as $\kappa/\lambda=0.0198$. 
Considering the successful relation 
$$
\frac{m_d m_s}{m_b^2}\simeq 4 \frac{m_e m_\mu}{m_\tau^2} \ ,
\eqno(4.7)
$$ 
for $b_d\simeq -1$, we seek for the best fit point of
$ b_d=- e^{i\beta_d}$ ($\beta_d^2 \ll 1$).
The observed ratio $m_d/m_s$ fixes the value $\beta_d$ 
as $\beta_d=18^\circ$.  
Then we can obtain the reasonable quark mass ratios [4],
not only $m_i^u/m_j^u$, $m_i^d/m_j^d$,  but also 
$m_i^u/m_j^d$:
$$
\begin{array}{lll}
m_u=0.000234\ {\rm GeV}, & m_c=0.610\ {\rm GeV}, &
m_t=0.181\ {\rm GeV}, \\
m_d=0.000475\ {\rm GeV}, & m_s=0.0923\ {\rm GeV}, &
m_b=3.01\ {\rm GeV}. \\
\end{array} 
\eqno(4.8)
$$
Here, we have taken 
$(m_0\kappa/\lambda)_q /(m_0\kappa/\lambda)_e=3.02$
in order to fit the observed quark mass values at 
$\mu=m_Z$ [10]
$$
\begin{array}{lll}
m_u=0.000233\ {\rm GeV}, & m_c=0.677\ {\rm GeV}, &
m_t=0.181\ {\rm GeV}, \\
{\ \ \ \ \ \ }_{-0.000045}^{+0.000042} &
{\ \ \ \ \ \ }_{-0.061}^{+0.056} & 
{\ \ \ \ \ \ }\pm 13 \\
m_d=0.000469\ {\rm GeV}, & m_s=0.0934\ {\rm GeV}, &
m_b=3.00\ {\rm GeV}. \\
{\ \ \ \ \ \ }_{-0.000066}^{+0.000060} &
{\ \ \ \ \ \ }_{-0.0130}^{+0.0118} &
{\ \ \ \ \ \ }\pm 0.11
\end{array} 
\eqno(4.9)
$$
We also obtain the reasonable values of the 
Cabibbo-Kobayashi-Maskawa (CKM) [11] matrix parameters:
$$
\begin{array}{ll}
|V_{us}|=0.220 \ , \ \ \ &  |V_{cb}|=0.0598 \ , \\ 
|V_{ub}|=0.00330 \ , \ \ \ & |V_{td}|=0.0155 \ . 
\end{array} \eqno(4.10)
$$
(The value of $|V_{cb}|$ is somewhat larger than the observed value.
For the improvement of the numerical value, see Ref.[12].)

So far, we have not mentioned why we call the present model (4.2)
``democratic" seesaw mass matrix model.
As far as the masses are concerned, the model with the democratic form
$$
M_F = \lambda m_0 \left[ \left( 
\begin{array}{ccc}
1 & 0 & 0 \\
0 & 1 & 0 \\
0 & 0 & 1 
\end{array} \right) + 
b_f \left( 
\begin{array}{ccc}
1 & 1 & 1 \\
1 & 1 & 1 \\
1 & 1 & 1 
\end{array} \right) \right] \ ,
\eqno(4.11)
$$
is equivalent to the model with the diagonal form
$$
M_F = \lambda m_0 \left( 
\begin{array}{ccc}
1 & 0 & 0 \\
0 & 1 & 0 \\
0 & 0 & 1+3b_f 
\end{array} \right) \ .
\eqno(4.12)
$$
However, for the prediction of the CKM matrix parameters,
the models show different features:
In the former model where the matrix $Z$ is diagonal, 
phases $\delta_i^f$ are brought 
into  the model as
$$
H_{mass}= y_L v_L \sum_i z_i \left( e^{i\delta_i^u}\, 
\overline{u}_{Li}u_{Ri} + 
 e^{i\delta_i^d}\, \overline{d}_{Li}d_{Ri}
 \right) + \cdots \ ,
\eqno(4.13)
$$
i.e.,
$$
Z_u=P(\delta_u) Z \ ,  \ \ \ \ Z_d =P(\delta_d) Z \ ,
\eqno(4.14)
$$
$$
P(\delta)\equiv {\rm diag}(e^{i\delta_1}, e^{i\delta_2}, 
e^{i\delta_3}) \ .
\eqno(4.15)
$$
Then, the observed CKM matrix parameters was 
successfully given [5] when we took
$$
P(\delta_u) P(\delta_d)^\dagger ={\rm diag}(1,\ 1,\ -1) \ .
\eqno(4.16)
$$
On the other hand, if we want the similar results for the 
latter case, we need a complicated form of the matrices $Z_u$
and $Z_d$:
$$
Z_u=A P(\delta_u) Z A^T \ ,  \ \ \ \ Z_d =A P(\delta_d) ZA^T \ ,
\eqno(4.17)
$$
where
$$
A=\left( 
\begin{array}{ccc}
\displaystyle\frac{1}{\sqrt{2}} & 
-\displaystyle\frac{1}{\sqrt{2}} & 0 \\
\displaystyle\frac{1}{\sqrt{6}} & 
\displaystyle\frac{1}{\sqrt{6}} & 
-\displaystyle\frac{2}{\sqrt{6}} \\
\displaystyle\frac{1}{\sqrt{3}} & 
\displaystyle\frac{1}{\sqrt{3}} & 
\displaystyle\frac{1}{\sqrt{3}} 
\end{array} \right) \ .
\eqno(4.18)
$$
Because of the simplicity of the former model, we consider that
the democratic basis of $M_F$ has a deep meaning.

\vspace{.2in}

{\large\bf 5. Application to the neutrino mass matrix}

\vglue.05in

The model can readily give a large mixing between two
neutrino states by taking $b_\nu\simeq -1/2$ or
$b_\nu\simeq -1$ as anticipated from Fig.~2. 
For example, the choice of $b_\nu\simeq -1/2$ gives 
$$
m_1^\nu \ll m_2^\nu \simeq m_3^\nu \ , \eqno(5.1)
$$
and
$$
U_{L} \simeq \left(
\begin{array}{ccc}
1 & \frac{1}{\sqrt{2}}\left(\sqrt{\frac{m_e}{m_\mu}}- 
\sqrt{\frac{m_e}{m_\tau}}\right) 
& \frac{1}{\sqrt{2}}\left(\sqrt{\frac{m_e}{m_\mu}}+ 
\sqrt{\frac{m_e}{m_\tau}}\right)  \\
-\sqrt{m_e/m_\mu} & \frac{1}{\sqrt{2}} & 
-\frac{1}{\sqrt{2}} \\
-\sqrt{m_e/m_\tau} & \frac{1}{\sqrt{2}} & 
\frac{1}{\sqrt{2}} \\
\end{array} \right) \ .\eqno(5.2)
$$
This is favorable to the large mixing picture 
suggested by the atmospheric neutrino data [13]. 
However, in order to give the simultaneous 
explanation of the atmospheric and solar neutrino [14]
data, we need a further study.

Examples based on the scenario A and scenario B are 
found in Refs.~[15] and [8], respectively.

\vspace{.2in}

{\large\bf 6. Concluding remarks}

\vglue.05in

In conclusion, we have pointed out the following 
features of the seesaw mass matrix model:

\noindent (i) The seesaw mass matrix  with 
the form $M_F$=[(unit matrix)+(rank-one matrix)]
and det$M_U=0$ can naturally understand the observed 
facts $m_t \gg m_b$ in spite of $m_u \sim m_d$ and  
why $m_t \sim \Lambda_W$,

\noindent 
(ii) The democratic seesaw mass matrix model 
with the input $ b_e=0 $ 
can give reasonable quark mass ratios and CKM matrix 
by taking $b_u=-1/3$ and $b_d= - e^{i18^\circ}$, 
and a large neutrino mixing $\nu_\mu$-$\nu_\tau$ 
by taking $b_\nu \simeq -1/2$.

\noindent
(iii) The model will provide new physics in abundance 
if $\Lambda_R \sim$ a few TeV:
we can expect observations of the fourth up-quark $t'$ with 
$m_{t'}\sim$ a few TeV and FCNC effects due to 
the abnormal structure of $U_R^u$.

However, this model is still in its beginning stages
and there are many future tasks:

(i) How do we understand the fermions $f$ and $F$?
Many ideas have been proposed for the unified 
understanding of the quarks and leptons $f$, while
in such a unification model 
there are no seats which should be assigned to 
the fermions $F_L$ and $F_R$. 

For this question, for example, we can understand
that the fermions $(f_L, \overline{F}_R)$ belong to 
{\bf 16} of a unification symmetry SO(10)$_L$ and 
$(f_R, \overline{F}_L)$ belong to {\bf 16} of
another unification symmetry  SO(10)$_R$, 
and the symmetries SO(10)$_L \times$SO(10)$_R$
are broken as follows:
$$ SO(10)_L \times SO(10)_R $$
$$ \Downarrow \ \ \Lambda_{GUT} $$
$$ [SU(3)\times SU(2)\times U(1)]_L\times 
 [SU(3)\times SU(2)\times U(1)]_R  $$
$$ \Downarrow \ \ \Lambda_{S} $$
$$ \ \ \ \ \ \ \ \ \ \ \ \ \ \ \ \ \ \ 
{\rm because\ of}\ \langle\overline{F}_L F_R\rangle $$
$$ SU(3)_c \times SU(2)_L \times SU(2)_R \times U(1)_Y $$
$$ \Downarrow \ \ \Lambda_{R} $$
$$ SU(3)_c \times SU(2)_L \times U(1)_Y $$
$$ \Downarrow \ \ \Lambda_{L} $$
$$ SU(3)_c \times U(1)_{em} $$

However, regrettably, we found [16] that the numerical results of 
the evolution of the gauge coupling constants conflicts
with the observed low energy data.

(ii) How do we understand the structure of $Z$?
In the  present stage, the values of $z_i$ have
been given by hand, i.e., by taking $b_e=0$.
For an  attempt to understand the structure of $Z$, for example, 
see Ref.~[17].

(iii) How do we understand the structure of $M_F$,
especially, the parameter $b_f$?
For example, there is a correlation between
the parameter $b_f$ and electric charge $Q$:
$$
\begin{array}{cccccccc}
 Q_\nu = 0\ ,& & Q_e = -1 \ ,& & Q_d = -1/3 \ ,& & Q_u=2/3 \ ,& \\
 &  \searrow & & \searrow & & \searrow &  &  \searrow \\
b_\nu =\ ? \ ,& & b_e =0 \ ,& & b_d\simeq -1 \ ,& &  b_u = -1/3 \ ,& \\
\end{array}
$$
Is it accidental?
At present, we have no idea.

We hope that many people direct their attention to 
the universal seesaw mass matrix model and 
thereby a great development of the quark and lepton
physics will be promoted along the line suggested by
present model.

\vspace*{.2in}
\centerline{\large\bf Acknowledgements}

\vglue.05in

The author would like to thank the organizers of this 
workshop for making the opportunity for my talk.
He also thank Prof.~H.~Fusaoka for his enjoyable and  
powerful collaboration.
This work was supported by the Grand-in-Aid for Scientific
Research, Ministry of Education, Science and Culture,
Japan (No.~08640386).

\vglue.2in
\newcounter{0000}
\centerline{\large\bf References}
\begin{list}
{[~\arabic{0000}~]}{\usecounter{0000}
\labelwidth=0.8cm\labelsep=.1cm\setlength{\leftmargin=0.7cm}
{\rightmargin=.2cm}}
\item For example, G.~C.~Branco, talk at this workshop.
\item Z.~G.~Berezhiani, Phys.~Lett.~{\bf 129B}, 99 (1983);
Phys.~Lett.~{\bf 150B}, 177 (1985);
D.~Chang and R.~N.~Mohapatra, Phys.~Rev.~Lett.~{\bf 58},1600 (1987); 
A.~Davidson and K.~C.~Wali, Phys.~Rev.~Lett.~{\bf 59}, 393 (1987);
S.~Rajpoot, Mod.~Phys.~Lett. {\bf A2}, 307 (1987); 
Phys.~Lett.~{\bf 191B}, 122 (1987); Phys.~Rev.~{\bf D36}, 1479 (1987);
K.~B.~Babu and R.~N.~Mohapatra, Phys.~Rev.~Lett.~{\bf 62}, 1079 (1989); 
Phys.~Rev. {\bf D41}, 1286 (1990); 
S.~Ranfone, Phys.~Rev.~{\bf D42}, 3819 (1990); 
A.~Davidson, S.~Ranfone and K.~C.~Wali, 
Phys.~Rev.~{\bf D41}, 208 (1990); 
I.~Sogami and T.~Shinohara, Prog.~Theor.~Phys.~{\bf 66}, 1031 (1991);
Phys.~Rev. {\bf D47}, 2905 (1993); 
Z.~G.~Berezhiani and R.~Rattazzi, Phys.~Lett.~{\bf B279}, 124 (1992);
P.~Cho, Phys.~Rev. {\bf D48}, 5331 (1994); 
A.~Davidson, L.~Michel, M.~L,~Sage and  K.~C.~Wali, 
Phys.~Rev.~{\bf D49}, 1378 (1994); 
W.~A.~Ponce, A.~Zepeda and R.~G.~Lozano, 
Phys.~Rev.~{\bf D49}, 4954 (1994).
\item M.~Gell-Mann, P.~Rammond and R.~Slansky, in {\it Supergravity}, 
edited by P.~van Nieuwenhuizen and D.~Z.~Freedman (North-Holland, 
1979); 
T.~Yanagida, in {\it Proc.~Workshop of the Unified Theory and 
Baryon Number in the Universe}, edited by A.~Sawada and A.~Sugamoto 
(KEK, 1979); 
R.~Mohapatra and G.~Senjanovic, Phys.~Rev.~Lett.~{\bf 44}, 912 (1980).
\item CDF Collaboration, F.~Abe {\it et al.}, Phys.~Rev.~Lett. 
{\bf 73}, 225 (1994).
\item Y.~Koide and H.~Fusaoka, Z.~Phys. {\bf C71}, 459 (1996). 
\item T.~Morozumi, T.~Satou, M.~N.~Rebelo and M.~Tanimoto, Phys.~Lett. 
{\bf B410}, 233 (1997).
\item Z.~G.~Berezhiani, in Ref.[2]; 
A.~Davidson and K.~C.~Wali, in Ref.[2];
S.~Rajpoot, in Ref.[4];
A.~Davidson, S.~Ranfone and K.~C.~Wali, in Ref.[2];
W.~A.~Ponce, A.~Zepeda and R.~G.~Lozano, in Ref.[2].
\item Y.~Koide, hep-ph/9707505 (1997), to be published in 
Phys.~Rev. {\bf D57}, No.7 (1998). 
\item Y.~Koide, Phys.~Rev. {\bf D56}, 2656 (1997).
\item H.~Fusaoka and Y.~Koide, Phys.~Rev. {\bf D57}, 3986 (1998).
\item N.~Cabibbo, Phys.~Rev.~Lett.~{\bf 10}, 531 (1996); 
M.~Kobayashi and T.~Maskawa, Prog.~Theor.~Phys.~{\bf 49}, 652 (1973).
\item Y.~Koide and H.~Fusaoka, 
Prog.~Theor.~Phys. {\bf 97}, 459 (1997).
\item Y.~Fukuda {\it et al.}, Phys.~Lett. {\bf B335}, 237 (1994).
\item For example, P.~Anselmann {\it et al.}, GALLEX collaboration, 
Phys.~Lett. {\bf B327}, 377 (1994); {\bf B357}, 237 (1995); 
J.~N.~Abdurashitov {\it et al.}, SAGE collaboration, {\it ibid.} 
{\bf B328}, 234 (1994). See also, N.~Hata and P.~Langacker, 
Phys.~Rev. {\bf D50}, 632 (1994); {\bf D52}, 420 (1995).
\item Y.~Koide, Mod.~Phys.~Lett. {\bf A36}, 2849 (1996).
\item Y.~Koide, in preparation.
\item Y.~Koide and M.~Tanimoto, Z.~Phys. {\bf C72}, 333 (1996);
A.~Davidson, T.~Schwartz and R.~R.~Volkas, hep-ph/9802235, (1998).
%
%
\end{list}

\end{document}